\documentclass[10pt,letterpaper]{article}
\usepackage[top=0.85in,left=2.75in,footskip=0.75in]{geometry}

\usepackage{amsmath,amssymb}

\usepackage{changepage}

\usepackage{textcomp,marvosym}

\usepackage{cite}

\usepackage{nameref,hyperref}

\usepackage[right]{lineno}

\usepackage[nopatch=eqnum]{microtype}
\DisableLigatures[f]{encoding = *, family = * }

\usepackage[table]{xcolor}

\usepackage{array}

\newcolumntype{+}{!{\vrule width 2pt}}

\newlength\savedwidth

\raggedright
\usepackage[aboveskip=1pt,labelfont=bf,labelsep=period,justification=raggedright,singlelinecheck=off]{caption}

\makeatletter
\renewcommand{\@biblabel}[1]{\quad#1.}
\makeatother

\usepackage{lastpage,fancyhdr,graphicx}
\usepackage{epstopdf}
\fancyheadoffset[L]{2.25in}
\fancyfootoffset[L]{2.25in}
\begin{document}

\vspace*{0.2in}

\begin{flushleft}
{\Large
\textbf\newline{Mapping the Global Election Landscape on Social Media in 2024} 
}
\newline
\\
Giulio Pecile\textsuperscript{1\Yinyang},
Niccolò Di Marco\textsuperscript{1\Yinyang},
Matteo Cinelli\textsuperscript{1},
Walter Quattrociocchi \textsuperscript{1\ddag*},
\\
\bigskip
\textbf{1} Department of Computer Science, Sapienza University of Rome, Viale Regina Elena 295, Rome, Italy
\bigskip

%
%
\Yinyang These authors contributed equally to this work.

\ddag Corrensponding author



* walter.quattrociocchi@uniroma1.it

\end{flushleft}
\section*{Abstract}

In 2024, half of the global population is expected to participate in elections, offering researchers a unique opportunity to study online information diffusion and user behavior. This study investigates the media landscape on social media by analyzing Facebook posts from national political parties and major news agencies across Europe, Mexico, and India. Our methodology identifies key topics and evaluates public interaction, reflecting broader trends in political engagement. Using Principal Component Analysis, we distil these topics to uncover patterns of correlation and differentiation. This approach reveals dominant themes that engage global audiences, providing critical insights into the interplay between public opinion and digital narratives during a major electoral cycle. 
Our findings highlight how different topics resonate across political spectrums, shaping political debate and offering a comprehensive view of the interaction between media content, political ideology, and audience engagement.



\section*{Introduction}
The advent and proliferation of social media platforms have fundamentally altered the information landscape and how we interact online \cite{avalle2024persistent}, transforming these platforms into essential tools for information \cite{pentina2014information, levy2021social}, entertainment \cite{etta2023characterizing,dimarco2024}, and personal communication \cite{boase2008personal}. As these platforms have become integrated into our daily lives, they have also merged entertainment-driven business models with complex social dynamics \cite{valensise2023drivers}, raising significant concerns about their impact on social dynamics, particularly in terms of polarization \cite{del2016echo, allcott2024effects, guess2023social, Falkenberg2022}, where content tailored to reinforce existing beliefs may eventually end up fostering societal division \cite{cinelli2021echo} and misinformation spreading \cite{del2016spreading,lazer2018science,van2022misinformation}.

Thus, social media platforms have revealed to be critical arenas which saw an explosion of information and misinformation for global events, including the COVID-19 pandemic \cite{cinelli2020covid, briand2021infodemics}, political events \cite{etta2024topology,flamino2023political,metaxas2012social}  and discussions on emerging technologies such as large language models \cite{alipour2024cross,luceri2024leveraging} and their implications for communication and automation. 
Moreover, these platforms play a pivotal role in consolidating biased views of the political landscape, potentially influencing public opinion and voter behavior during elections \cite{bovet2019influence, flamino2023political} through the rapid dissemination and amplification of political content \cite{pennycook2021psychology}.

Research on how platform-specific effects influence social dynamics has highlighted several challenges \cite{cinelli2021echo, gonzalez2023social, guess2023social, nyhan2023nature, guess2023reshares}. 
Indeed, these interactions often exhibit persistent patterns despite different platforms, topics, and contexts, suggesting underlying consistencies in online human behavior \cite{avalle2024persistent}. Online users show the tendency to selectively expose to information \cite{cinelli2020selective,bakshy2015exposure}, preferring content that aligns with their pre-existing beliefs while avoiding contrary evidence \cite{bessi2015science, zollo2017debunking}. This behavior may fosters the emergence of homophilic communities, also know as echo-chambers \cite{del2016echo}, which significantly influence belief formation and communication methods \cite{cinelli2021echo}, especially during delicate periods such as elections. 

As the 2024 elections approach, with a substantial global population expected to vote, Facebook is a crucial platform for electoral campaigning. This period offers an invaluable opportunity to comprehensively analyze how different countries use social media for news consumption, public debate, and electoral influence.

In this study, we analyze content from news agencies and political parties on Facebook to see how it affects user interactions (including reactions, comments and shares) and examine variations between countries and the political leanings of parties. Using Principal Component Analysis, our analysis shows the main trends of the public discourse within the social media ecosystem, allowing us to identify patterns of both correlation and differences among key themes based on geographic and political distinctions.

Our comparative study across multiple countries aims to provide a multilateral understanding of how discussions on social media reflect electoral dynamics in a pivotal election year. Additionally, this paper explores how different countries, and thus socio-cultural factors, shape user interactions and interests on social media, contributing to the global dialogue on democracy and public discourse.

The structure of the paper is the following: We first assess the level of engagement on social media across various countries, then we explore the controversy surrounding discussions on diverse topics and, finally, we categorize the information and topics based on political leanings, providing a comprehensive overview of the digital electoral landscape in 2024.

\section*{Materials and methods}

\subsection*{Dataset}

For our analysis, we manually compiled detailed lists of Facebook pages from each country in the European Union, as well as the UK, India, and Mexico. These lists include pages linked to the websites of political parties represented in national parliaments and major news agencies in each country. We also included major American news agencies and public Facebook pages of influential politicians, identified using the YouGov public survey platform \cite{yougov}. Our dataset includes 508 news agencies and 336 political parties across 31 countries, capturing a total of approximately 4.2 million and 176 thousand posts between September 1, 2023, and May 1, 2024. For each post, we recorded the page name, posting time, textual content, and aggregated metrics like the number of reactions and comments.

\subsection*{News outlet and parties labeling}
In our study, we analyze Facebook pages from various countries that provide diverse perspectives on current events, ranging from the progressive left to the conservative right. We manually check political parties' ideological leanings, since they are often clear and aligned with international coalitions, such as those within the European Parliament. Unlike political parties, news outlets usually have less explicit political alignments. To classify these, we use ratings from Media Bias/Fact Check (MBFC), an independent website that evaluates news agencies based on factuality and political bias.
While MBFC recognizes a wide range of political leanings, for analytical simplicity, we consolidate these into three primary categories: `Left,' `Center,' and `Right.' 

\subsection*{Process of Topic Modelling}
We implemented a two-step process to assess the topics discussed in each post. We processed each country separately since each one had posts in their primary language. For each country, we selected a random sample of 50,000 posts made by news organizations. We then cleaned the text of these posts by removing links, hashtags, and non-alphanumeric characters, such as emojis. Subsequently, we applied the BERTopic clustering algorithm to perform topic clustering for each country. The resulting clusters were labelled with overarching themes such as `economy,' `crime,' `entertainment,' and ongoing conflicts in Ukraine and the Middle East (identified as `mena' in the analysis).
Notice that, Finland and Sweden did not produce any discernible topics during the clustering phase.
To infer the topic of all posts, we computed the term frequency-inverse document frequency (tf-idf) values of the words from the posts within each cluster, treating the concatenation of all posts from each macro-topic as a single document. This allowed us to identify the macro-topic that most closely aligned with the textual content of each post based on the tf-idf scores of the words.

\begin{figure}[!h]
    \caption{{\bf Topics discussed as share of total classified posts.} The plurality of posts from political parties is about internal politics.}
    \includegraphics[width = \linewidth]{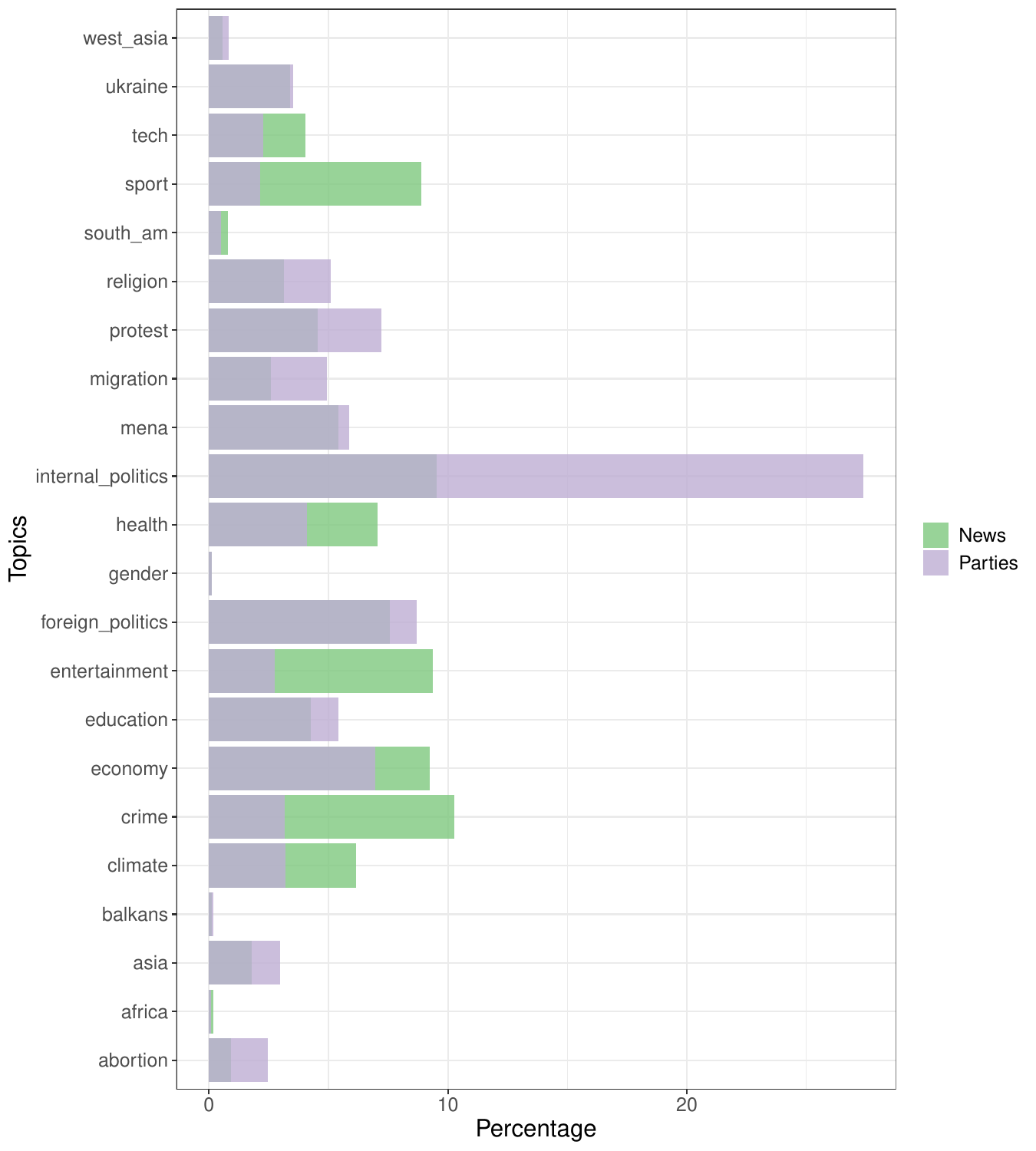}
    \label{fig:post_hist_perc}
\end{figure}

Fig.  \ref{fig:post_hist_perc} presents the share of posts discussing each topic. As shown, political parties predominantly post about internal politics, whereas `lighter' topics such as entertainment or sports garner significantly more attention from news outlets.

\subsection*{Principal component analysis}

Principal Component Analysis (PCA) \cite{pca} is a method for linear dimensionality reduction, widely used to summarize and visualize the information of datasets containing a large number of individuals or observations in lower dimensions containing the greater possible variance \cite{Jolliffe2016-rf,Jolliffe2002-fy}.

In this work, we mostly use biplots \cite{GABRIEL1971} and methods to visualize the overall relation of individuals or variables. Therefore we briefly recall some aspects used in this work. 

Consider a matrix $M$ of dimension $n \times m$ in which rows are individuals and columns are variables measured on each individual. The entry $M_{i,j}$ contains the $j-$th variable of individual $i$. 

After applying PCA to the columns of $M$ (each column is a variable) it is common to represent the results in a biplot, a two-dimensional space containing the larger possible variance for both variables and individuals, even if their coordinates are not constructed onto the same space. 

Here, we summarize their properties:

\begin{itemize}
    \item the distance between the variables and the origin is a measure of how well that variable is explained in two dimensions;
    \item if $i$ and $j$ are variables, we have $\cos\theta_{ij} = r_{ij}$, i.e. the cos of the angle between two variables is equal to their correlation. Therefore, positively correlated variables are grouped together, while negatively correlated variables are positioned on opposite quadrants;
    \item individuals with similar values of variables are grouped together;
    \item an individual that is on the same (opposite) side of a certain variable has a high (low) value for that variable.
\end{itemize}

Note that variables and individuals could also be represented separately, keeping the properties listed above.

In this work, we use the packages {\it FactoMineR} \cite{FactomineR} and {\it factoextra} \cite{factoextra} to apply PCA and construct the relative biplots. 
In particular, we represent each individual as a black point, while variables are represented using red vectors starting from the origin. 



\section*{Results and Discussion}


\subsection*{Engagement across countries}
\begin{figure}[!h] 
\caption{{\bf Distribution of reactions.} $(a)$ News outlets dataset and $(b)$ Parties dataset.} 
\includegraphics[width = \linewidth]{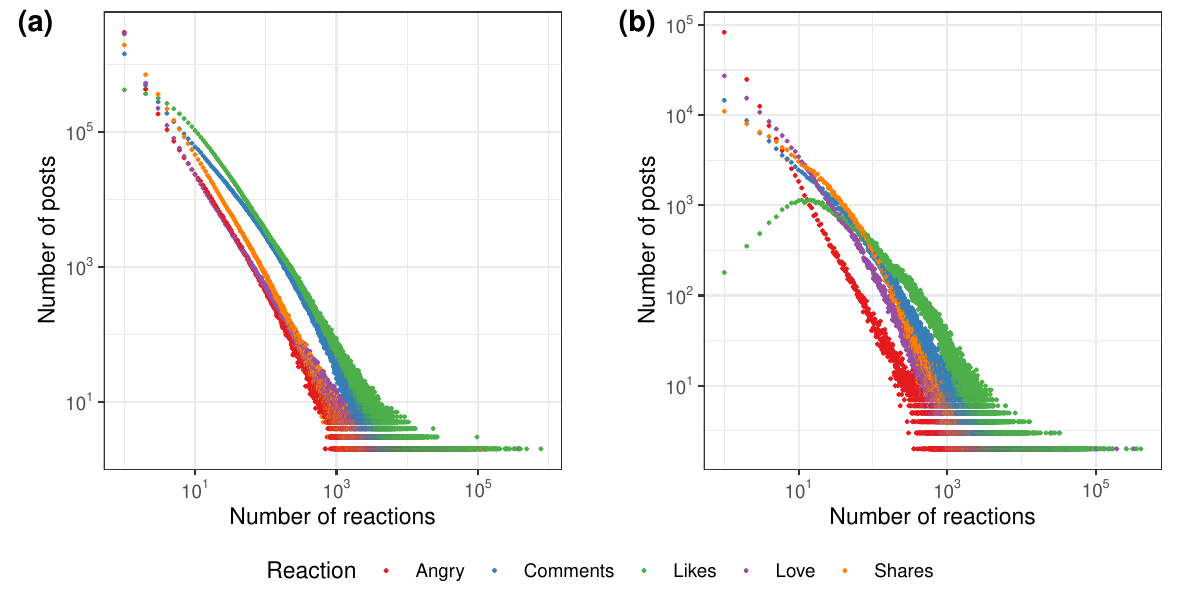} 
\label{fig} \end{figure}

We start the analysis by exploring the distribution of user engagement, which we measure through the reactions received by posts. Fig. \ref{fig} illustrates the distribution of these interactions for both news outlets and political parties.

We observe that reactions across different types of content exhibit similar distributions. This pattern holds true for both news agencies and political parties, as illustrated in Fig. \ref{fig} (a) and (b). However, a significant deviation occurs in the likes received by political parties, where the mode of the distribution is approximately $10$.  This indicates a reduced frequency of posts that receive very few likes for these entities. 
Moreover, consistent with prior research \cite{avalle2024persistent}, heavy tails characterize all the observed distributions, suggesting that while most posts receive relatively few reactions, a small number accumulates a disproportionately large number.

Interestingly, the number of reactions received by posts from political parties and news agencies falls within similar orders of magnitude, with tails ranging between $10^3$ and $10^5$ reactions.
Engagement distribution patterns for each country are displayed in Fig. \ref{fig:engagement_si_n} for news agencies and \ref{fig:engagement_si_p} for political parties, exhibiting the same consistent patterns.

\subsection*{Controversial topics across countries}
\begin{figure}[!h]
    \caption{\textbf{Boxplots of love-angry scores received for each topic}}
    \label{fig:boxall}
    \centering
    \includegraphics[width=\linewidth]{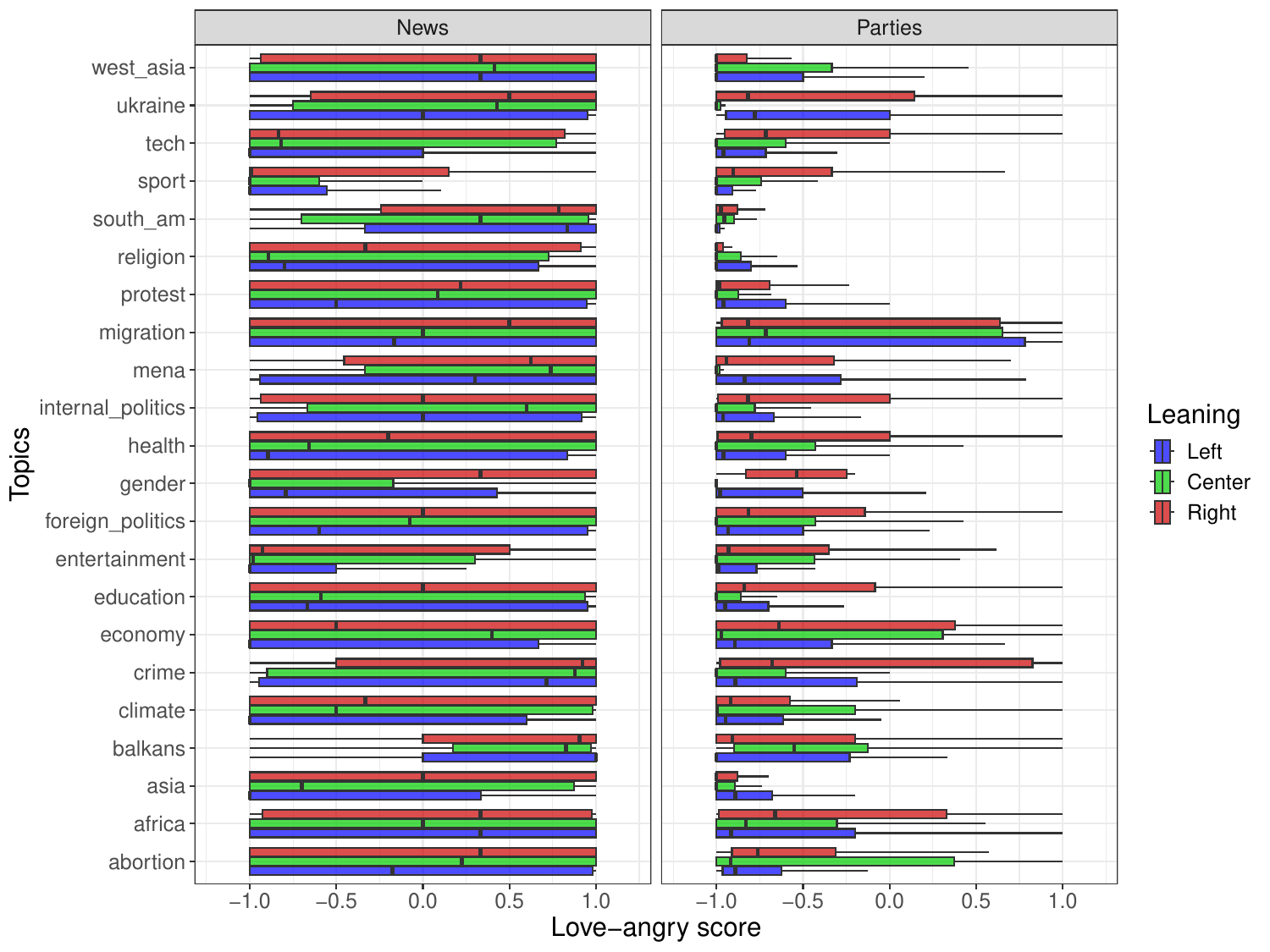}
\end{figure}

In this section, we investigate which topics capture more contentious debate by utilizing a metric derived from the number of `love' and `angry' reactions each post receives. Specifically, we define the \textit{love-angry} score \cite{etta2023characterizing} for post $i$, denoted as $LA(i)$, using the following equation:

\begin{equation*}
    LA(i) = \frac{a_i - l_i}{a_i + l_i},
\end{equation*}

where $a_i$ represents the number of angry reactions and $l_i$ the number of love reactions received by post $i$. This metric serves to quantify the degree of divisiveness elicited by each post. In particular, the score takes values in $[-1, +1]$, with $-1$ ($+1$) corresponding to posts that have received only love (angry) reactions. 
Fig. \ref{fig:boxall} displays the distribution of love-angry scores of news agencies and political parties grouped according to their political leaning. 

The results show that news agencies often receive higher `love-angry' scores on their posts than political parties. This suggests news agencies reach a wider audience, attracting diverse reactions. On the other hand, posts from political parties generally see lower scores, indicating they might be engaging a more specific group of followers that are likely to share similar views and which could imply less debate and more agreement, typical of an echo chamber effect \cite{cinelli2021echo,del2016echo}.

\subsection*{Topics in each country}
In this section, we aim to highlight the most discussed topics in each country, using the total number of interactions as a proxy for their relative perceived importance. 

We define $C$ as the number of distinct countries and $T$ as the number of topics discussed. To mitigate potential biases due to the varying numbers of news outlets across countries, we construct a matrix $M$ of dimensions $C \times T$, where the entry $M_{ij}$ represents the fraction of total engagement received in topic $j$ by country $i$. 

\begin{figure}[!h]
    \caption{\textbf{Most engaged topics in each country, as revealed by PCA.}}
    \label{fig:pca_countries_topics}
    \centering
    \includegraphics[width=\linewidth]{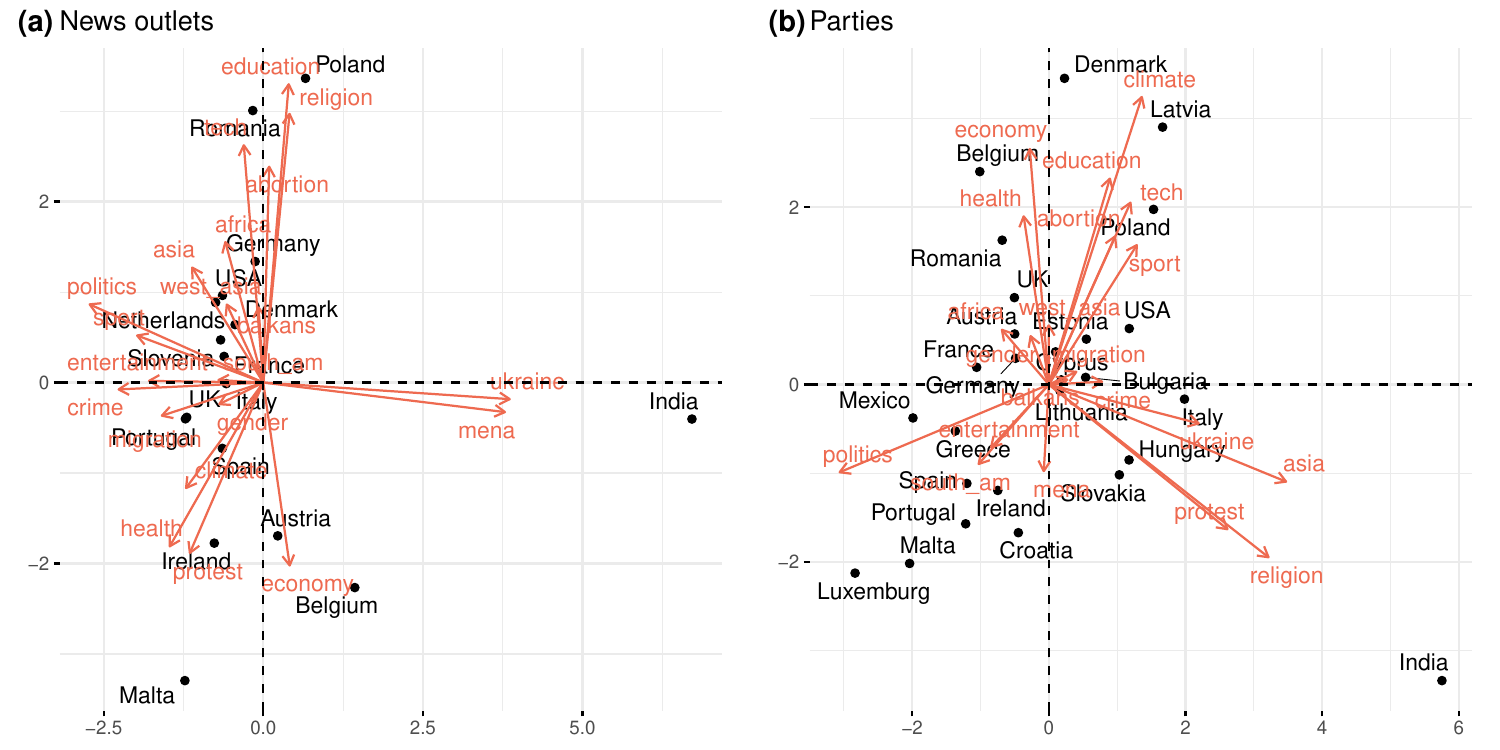}

\end{figure}

To distinguish between the discussions led by news outlets and political parties, we create two separate matrices, $M^{news}$ and $M^{parties}$. Principal Component Analysis (PCA) is then applied to these matrices to derive a two-dimensional representation that explains the most variance; further details can be found in the Materials and Methods section.
Fig. \ref{fig:pca_countries_topics} presents the biplots from $M^{news}$ and $M^{parties}$.

 Notably, news outlets in India demonstrate unusually high engagement compared with other countries on topics such as the Middle East (abbreviated as MENA) and Ukraine. For most other countries, engagement patterns on the first principal component, which captures the largest variance, are relatively uniform, suggesting similar interest levels in entertainment, crime, and politics.

However, specific topics such as education, religion, technology, abortion, and africa show an inverse relationship with others like climate, health, protest, and economy. Countries like Ireland, Austria, Malta, and Spain are more engaged with the latter topics while less so with the former.
From the perspective of political parties, engagement patterns vary significantly. For instance, parties in Northern European countries like Denmark, Belgium, and Latvia actively engage more with climate change, economy, education, health, abortion, technology, and sports.
In contrast, topics such as protest, asia, religion, and ukraine obtain higher interactions in countries like Italy, India, Hungary, and Slovakia. This highlights a clear divergence in the focus of discussions between news outlets and political entities.

\subsection*{Topics and political leaning}
In the previous section, we extracted the most engaging topics from each country without considering the specific landscape of news outlets and political parties within those countries. In this section, we address this gap by applying a similar analysis distinctly for each nation, incorporating an additional dimension of political leaning (details are provided in the Materials and Methods section).

\begin{figure}[!h]
    \caption{\textbf{Biplot of news outlets grouped by engaged topics for countries with $n_c \geq 3$.}}
    \label{fig:pca_news_topics}
    \centering
    \includegraphics[width=\linewidth]{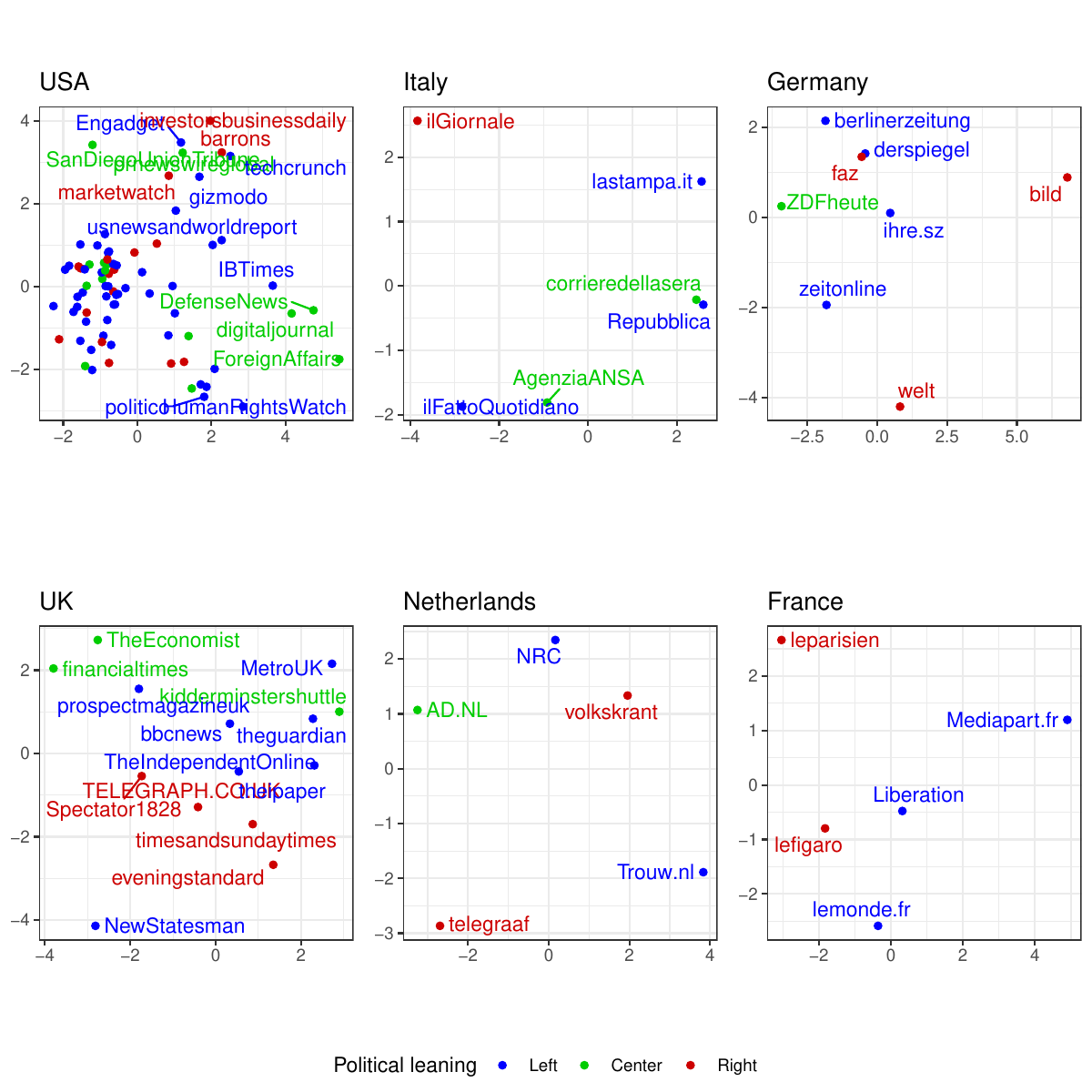}
\end{figure}

We focus on the news outlets case (the parties are treated similarly). Following a method similar to the previous one, we consider a country $c$ with $n_c$ labeled news outlets discussing $m_c$ topics. We construct the matrix $M^c$ of dimension $n_c \times m_c$, where $M^c_{ij}$ represents the fraction of engagement received on topic $j$ by news outlet $i$. Fig. \ref{fig:pca_news_topics} shows the biplot resulting after applying PCA to $M^c$ on each country having $n_c \geq 3$.

We conduct a similar analysis for political parties, whose results are displayed in Fig. \ref{fig:pca_parties_topics} for all countries having $n_c \geq 3$.

Interestingly, the analysis reveals that the political leaning of each entity does not significantly influence how engagement is distributed across topics.

One possible explanation for the observed engagement patterns can be attributed to the overall narrative style of each page. Even when discussing the same topics, different pages may present perspectives that resonate more closely with specific audiences. However, due to the limitations of our classification approach, PCA primarily captures the general content themes without distinguishing between nuanced perspectives.

Despite this limitation and the potential imbalances in labeling across different countries, our findings suggest that news outlets and political parties with also different political leaning —left, center, or right— are likely to engage with similar topics.

\begin{figure}[!h]
    \centering
    \includegraphics[width=.9\linewidth]{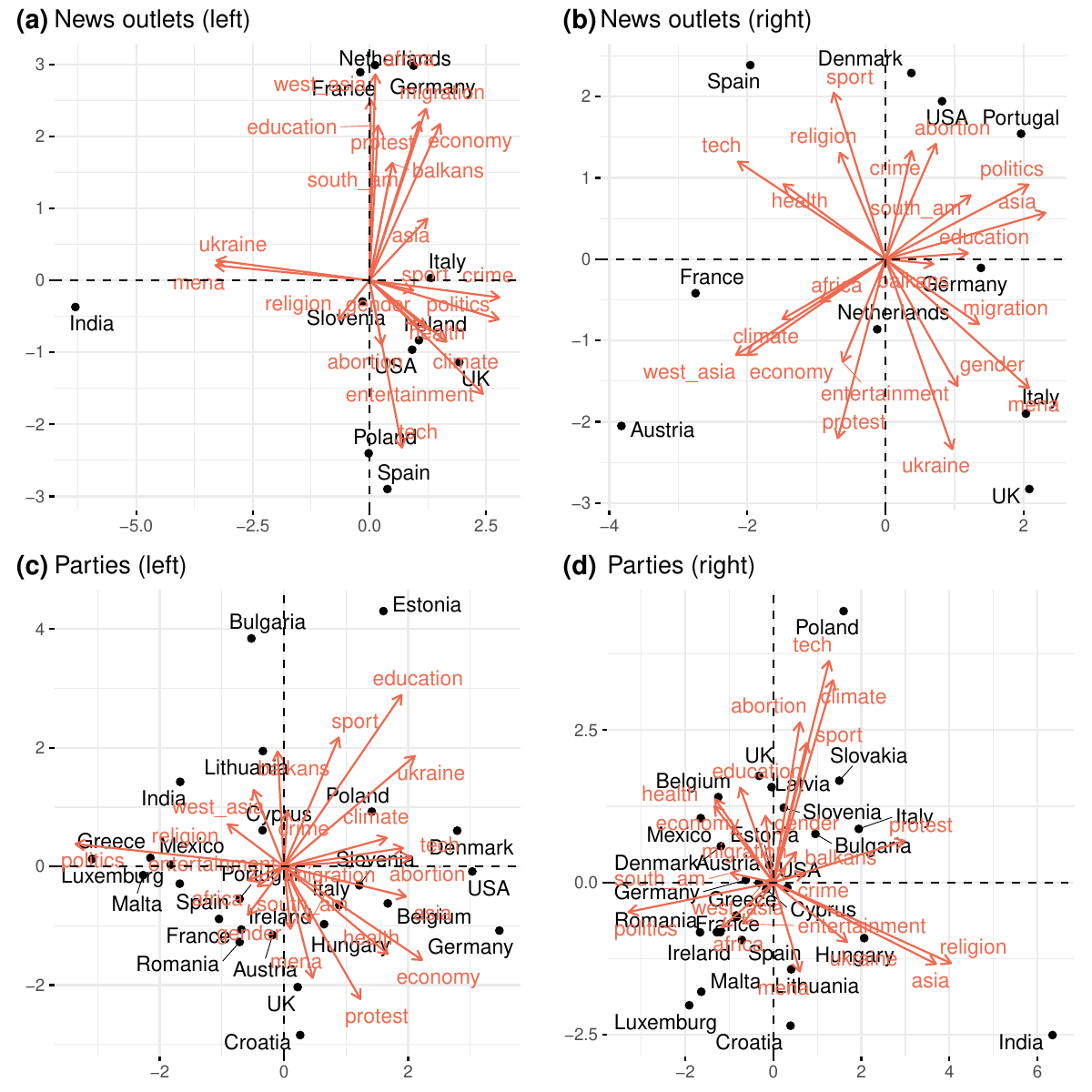}
    \caption{\textbf{Most engaging topics of news outlets and parties if only a specific leaning is considered.}}
    \label{fig:pca_leanings}
\end{figure}

To delve deeper into the engagement dynamics specific to each political leaning, we employ a similar analytical procedure. 
In particular, we construct a matrix $M^b$ in which the rows are countries and columns are topics. Differently from before, the generic entry $M^b _{i,j}$ is the fraction of engagement received by news outlets of country $i$, having political leaning $b \in \{left, right\}$, in topic $j$. PCA is then applied to these matrices to yield a two-dimensional representation of the topic discussed by a specific political leaning in each country.
The results are reported in Fig. \ref{fig:pca_leanings}.

We observe that left news outlets in Poland, Spain, USA, Ireland and UK are getting high engagement in topics such as abortion, religion, climate and tech. 
At the same time Netherlands, France and Germany are more focused on education, economy, africa and migration. 
On the other hand, right news outlet in Italy, Germany, UK and Netherlands are getting high engagement in topics such as gender, migration and ukraine. Interestingly, USA, Portugal, Denmark and Spain are more interested in religion, abortion, sport and crime. Finally, only right news outlets from France and Austria seem to get engagement with the climate topic.

Fig. \ref{fig:pca_leanings} $(c)-(d)$ depict the results of the same analysis for Political Parties. In this case we observe even more differences between each country, suggesting that political parties from different countries, even if with the same political leaning - left or right - tend to engage in varying and different topics. 

Finally, we are interested in highlighting the topics discussed by each political leaning in each country. To hit this aim, we employ the following procedure: For each country $c$, we construct a matrix $M^c$ of dimension $3 \times m_c$ where the rows represent the political leanings of the news outlets—\textit{left, center}, and \textit{right}. The entry $M^c_{ij}$ denotes the fraction of engagement received by news outlets with political leaning $i$ in topic $j$. PCA is then applied to this matrix to yield a two-dimensional representation capturing the most significant aspects of the data.

\begin{figure}[!h]
    \centering
    \includegraphics[width=.9\linewidth]{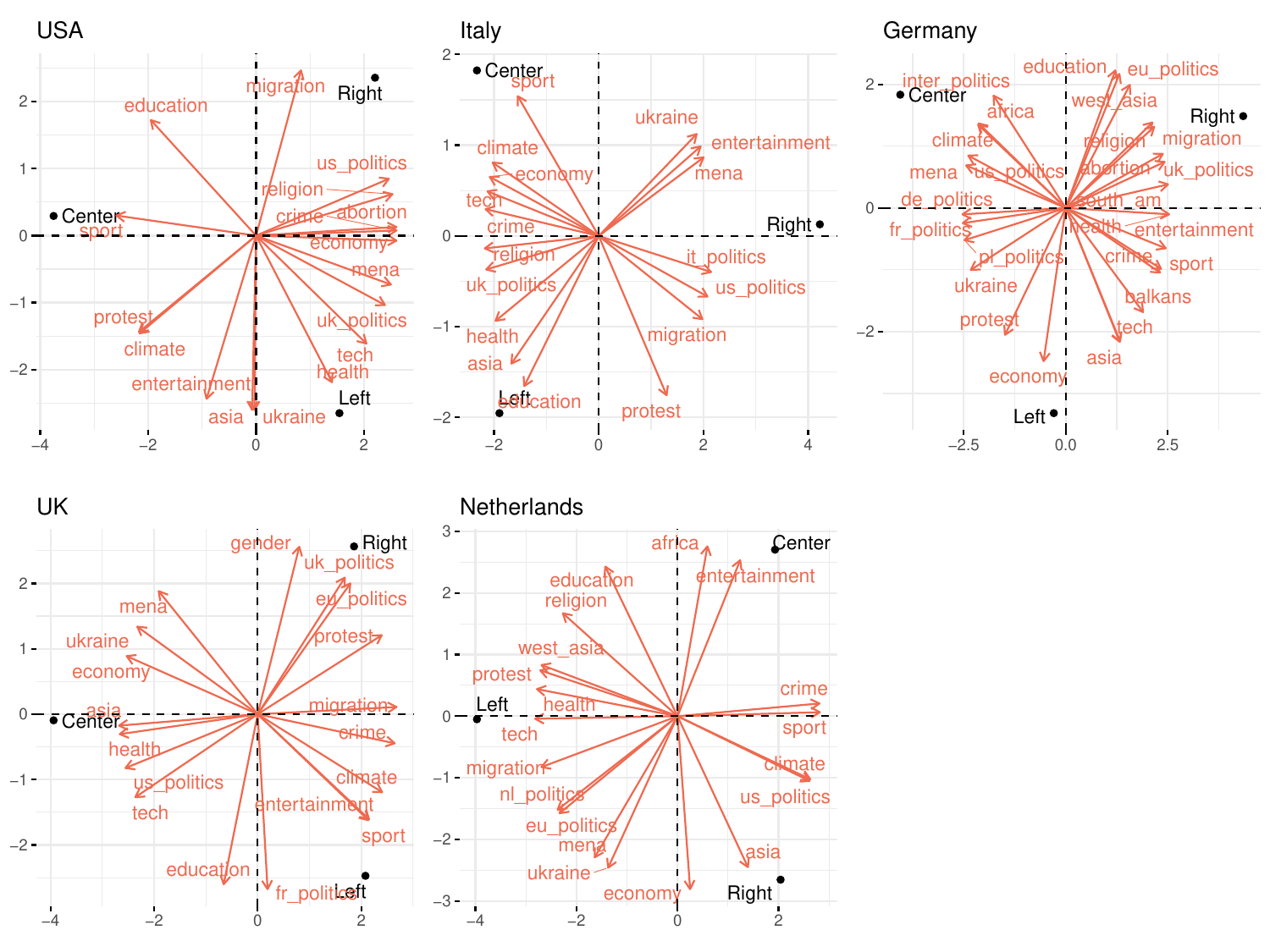}
    \caption{\textbf{Most engaging topics of news outlets by political leaning.}}
    \label{fig:pca_news_leanings}
\end{figure}

The resulting biplots generally indicate that right-leaning news outlets attract more engagement on topics related to politics, religion, and migration, whereas left-leaning news outlets are more actively engaged in topics like education, health, and technology. Additionally, the angles between topics suggest a slight negative correlation between the topics favored by the left and right, indicating a moderate level of, overall, topic segregation.

\begin{figure}[!h]
    \centering
    \includegraphics[width=.9\linewidth]{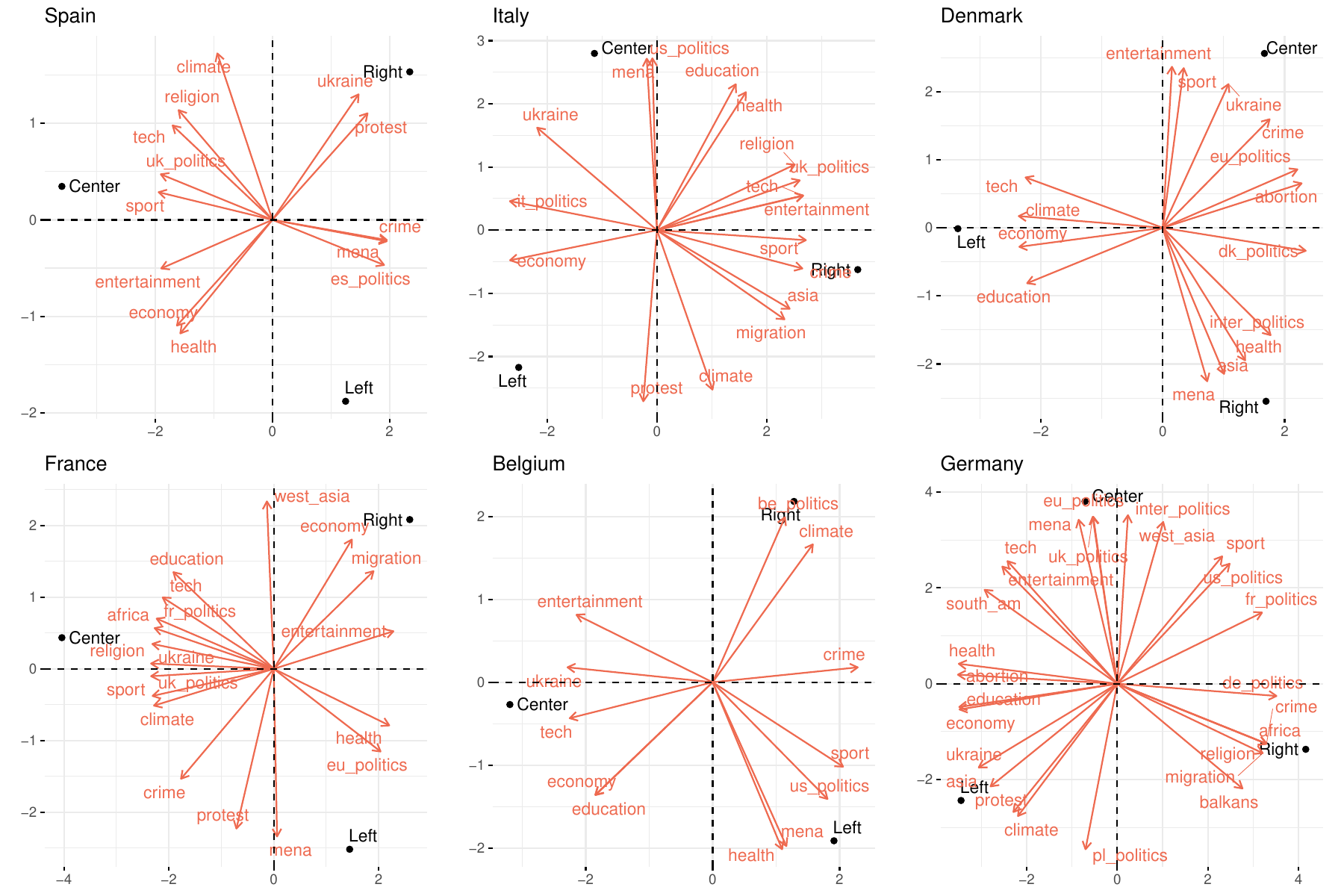}
    \caption{\textbf{Most engaging topics of parties by political leaning.}}
    \label{fig:pca_parties_leanings}
\end{figure}

Similarly, Fig. \ref{fig:pca_parties_leanings} displays the engagement patterns by political leaning for a subset of countries in the parties dataset. The complete results are detailed in Fig. \ref{fig:pca_parties_leanings_si}.

\section*{Conclusion}
This study analyzes how different political leanings and media outlets engage audiences on social media, revealing patterns in how various topics resonate with people globally. Despite challenges with data classification and PCA's limitations, our findings show clear trends in topic engagement that go beyond political lines.
Our analysis indicates that while news outlets and political parties often discuss similar broad topics, the intensity and nature of this engagement differ based on political orientation. Specifically, right-leaning outlets and parties focus more on politics, religion, and migration. In contrast, left-leaning groups emphasize education, health, and technology. This suggests that the left may be talking about things that need to capture more people's interest than the topics the right focuses on.
The PCA biplots show differences in topic engagement, but there isn't a sharp divide between the left and right. This means that even politically polarized groups can find common ground on specific issues from different perspectives. For instance, both sides might discuss the economy but approach it from their unique viewpoints.
These findings are important for policymakers, media strategists, and researchers. Understanding these engagement patterns can create more effective messages that appeal to a wider audience, bridging divides and fostering better public discourse. This study also offers practical insights for designing social media campaigns considering ideological differences and audience preferences.

\section*{Acknowledgments}
The work is supported by IRIS Infodemic Coalition (UK government, grant no. SCH-00001-3391), 
SERICS (PE00000014) under the NRRP MUR program funded by the European Union - NextGenerationEU, project CRESP from the Italian Ministry of Health under the program CCM 2022, PON project “Ricerca e Innovazione” 2014-2020, and PRIN Project MUSMA for Italian Ministry of University and Research (MUR) through the PRIN 2022. 
This work was supported by the PRIN 2022 “MUSMA” - CUP G53D23002930006 - Funded by EU - Next-Generation EU – M4 C2 I1.1.

\nolinenumbers
\bibliography{bibliography} 

\begin{thebibliography}{10}

\bibitem{avalle2024persistent}
Avalle M, Di~Marco N, Etta G, Sangiorgio E, Alipour S, Bonetti A, et~al.
\newblock Persistent interaction patterns across social media platforms and over time.
\newblock Nature. 2024;628(8008):582--589.

\bibitem{pentina2014information}
Pentina I, Tarafdar M.
\newblock From “information” to “knowing”: Exploring the role of social media in contemporary news consumption.
\newblock Computers in human behavior. 2014;35:211--223.

\bibitem{levy2021social}
Levy R.
\newblock Social media, news consumption, and polarization: Evidence from a field experiment.
\newblock American economic review. 2021;111(3):831--870.

\bibitem{etta2023characterizing}
Etta G, Sangiorgio E, Di~Marco N, Avalle M, Scala A, Cinelli M, et~al.
\newblock Characterizing engagement dynamics across topics on Facebook.
\newblock Plos one. 2023;18(6):e0286150.

\bibitem{dimarco2024}
Di~Marco N, Cinelli M, Alipour S, Quattrociocchi W.
\newblock Users Volatility on Reddit and Voat.
\newblock IEEE Transactions on Computational Social Systems. 2024; p. 1--9.
\newblock doi:{10.1109/TCSS.2024.3379318}.

\bibitem{boase2008personal}
Boase J.
\newblock Personal networks and the personal communication system: Using multiple media to connect.
\newblock Information, communication \& society. 2008;11(4):490--508.

\bibitem{valensise2023drivers}
Valensise CM, Cinelli M, Quattrociocchi W.
\newblock The drivers of online polarization: Fitting models to data.
\newblock Information Sciences. 2023;642:119152.

\bibitem{del2016echo}
Del~Vicario M, Vivaldo G, Bessi A, Zollo F, Scala A, Caldarelli G, et~al.
\newblock Echo chambers: Emotional contagion and group polarization on facebook.
\newblock Scientific reports. 2016;6(1):37825.

\bibitem{allcott2024effects}
Allcott H, Gentzkow M, Mason W, Wilkins A, Barber{\'a} P, Brown T, et~al.
\newblock The effects of Facebook and Instagram on the 2020 election: A deactivation experiment.
\newblock Proceedings of the National Academy of Sciences. 2024;121(21):e2321584121.

\bibitem{guess2023social}
Guess AM, Malhotra N, Pan J, Barber{\'a} P, Allcott H, Brown T, et~al.
\newblock How do social media feed algorithms affect attitudes and behavior in an election campaign?
\newblock Science. 2023;381(6656):398--404.

\bibitem{Falkenberg2022}
Falkenberg M, Galeazzi A, Torricelli M, Di~Marco N, Larosa F, Sas M, et~al.
\newblock Growing polarization around climate change on social media.
\newblock Nature Climate Change. 2022;12(12):1114--1121.

\bibitem{cinelli2021echo}
Cinelli M, De~Francisci~Morales G, Galeazzi A, Quattrociocchi W, Starnini M.
\newblock The echo chamber effect on social media.
\newblock Proceedings of the National Academy of Sciences. 2021;118(9):e2023301118.

\bibitem{del2016spreading}
Del~Vicario M, Bessi A, Zollo F, Petroni F, Scala A, Caldarelli G, et~al.
\newblock The spreading of misinformation online.
\newblock Proceedings of the national academy of Sciences. 2016;113(3):554--559.

\bibitem{lazer2018science}
Lazer DM, Baum MA, Benkler Y, Berinsky AJ, Greenhill KM, Menczer F, et~al.
\newblock The science of fake news.
\newblock Science. 2018;359(6380):1094--1096.

\bibitem{van2022misinformation}
Van Der~Linden S.
\newblock Misinformation: susceptibility, spread, and interventions to immunize the public.
\newblock Nature medicine. 2022;28(3):460--467.

\bibitem{cinelli2020covid}
Cinelli M, Quattrociocchi W, Galeazzi A, Valensise CM, Brugnoli E, Schmidt AL, et~al.
\newblock The COVID-19 social media infodemic.
\newblock Scientific reports. 2020;10(1):1--10.

\bibitem{briand2021infodemics}
Briand SC, Cinelli M, Nguyen T, Lewis R, Prybylski D, Valensise CM, et~al.
\newblock Infodemics: A new challenge for public health.
\newblock Cell. 2021;184(25):6010--6014.

\bibitem{etta2024topology}
Etta G, Cinelli M, Di~Marco N, Avalle M, Panconesi A, Quattrociocchi W.
\newblock A Topology-Based Approach for Predicting Toxic Outcomes on Twitter and YouTube.
\newblock IEEE Transactions on Network Science and Engineering. 2024;.

\bibitem{flamino2023political}
Flamino J, Galeazzi A, Feldman S, Macy MW, Cross B, Zhou Z, et~al.
\newblock Political polarization of news media and influencers on Twitter in the 2016 and 2020 US presidential elections.
\newblock Nature Human Behaviour. 2023;7(6):904--916.

\bibitem{metaxas2012social}
Metaxas PT, Mustafaraj E.
\newblock Social media and the elections.
\newblock Science. 2012;338(6106):472--473.

\bibitem{alipour2024cross}
Alipour S, Galeazzi A, Sangiorgio E, Avalle M, Bojic L, Cinelli M, et~al.
\newblock Cross-platform social dynamics: an analysis of ChatGPT and COVID-19 vaccine conversations.
\newblock Scientific Reports. 2024;14(1):2789.

\bibitem{luceri2024leveraging}
Luceri L, Boniardi E, Ferrara E.
\newblock Leveraging Large Language Models to Detect Influence Campaigns on Social Media.
\newblock In: Companion Proceedings of the ACM on Web Conference 2024; 2024. p. 1459--1467.

\bibitem{bovet2019influence}
Bovet A, Makse HA.
\newblock Influence of fake news in Twitter during the 2016 US presidential election.
\newblock Nature communications. 2019;10(1):7.

\bibitem{pennycook2021psychology}
Pennycook G, Rand DG.
\newblock The psychology of fake news.
\newblock Trends in cognitive sciences. 2021;25(5):388--402.

\bibitem{gonzalez2023social}
Gonz{\'a}lez-Bail{\'o}n S, Lelkes Y.
\newblock Do social media undermine social cohesion? A critical review.
\newblock Social Issues and Policy Review. 2023;17(1):155--180.

\bibitem{nyhan2023nature}
Nyhan B, Settle J, Thorson E, et~al.
\newblock Like-minded sources on Facebook are prevalent but not polarizing.
\newblock Nature. 2023; p. 137--144.

\bibitem{guess2023reshares}
Guess AM, Malhotra N, Pan J, Barber{\'a} P, Allcott H, Brown T, et~al.
\newblock Reshares on social media amplify political news but do not detectably affect beliefs or opinions.
\newblock Science. 2023;381(6656):404--408.

\bibitem{cinelli2020selective}
Cinelli M, Brugnoli E, Schmidt AL, Zollo F, Quattrociocchi W, Scala A.
\newblock Selective exposure shapes the Facebook news diet.
\newblock PloS one. 2020;15(3):e0229129.

\bibitem{bakshy2015exposure}
Bakshy E, Messing S, Adamic LA.
\newblock Exposure to ideologically diverse news and opinion on Facebook.
\newblock Science. 2015;348(6239):1130--1132.

\bibitem{bessi2015science}
Bessi A, Coletto M, Davidescu GA, Scala A, Caldarelli G, Quattrociocchi W.
\newblock Science vs conspiracy: Collective narratives in the age of misinformation.
\newblock PloS one. 2015;10(2):e0118093.

\bibitem{zollo2017debunking}
Zollo F, Bessi A, Del~Vicario M, Scala A, Caldarelli G, Shekhtman L, et~al.
\newblock Debunking in a world of tribes.
\newblock PloS one. 2017;12(7):e0181821.

\bibitem{yougov}
The Most Popular Politicians;.
\newblock \url{https://today.yougov.com/ratings/politics/popularity/politicians/all}.

\bibitem{pca}
S KPFR.
\newblock LIII. On lines and planes of closest fit to systems of points in space.
\newblock The London, Edinburgh, and Dublin Philosophical Magazine and Journal of Science. 1901;2(11):559--572.
\newblock doi:{10.1080/14786440109462720}.

\bibitem{Jolliffe2016-rf}
Jolliffe IT, Cadima J.
\newblock Principal component analysis: a review and recent developments.
\newblock Philos Trans A Math Phys Eng Sci. 2016;374(2065):20150202.

\bibitem{Jolliffe2002-fy}
Jolliffe IT.
\newblock Principal Component Analysis.
\newblock 2nd ed. Springer Series in Statistics. New York, NY: Springer; 2002.

\bibitem{GABRIEL1971}
GABRIEL KR.
\newblock The biplot graphic display of matrices with application to principal component analysis.
\newblock Biometrika. 1971;58(3):453–467.
\newblock doi:{10.1093/biomet/58.3.453}.

\bibitem{FactomineR}
L\^e S, Josse J, Husson F.
\newblock {FactoMineR}: A Package for Multivariate Analysis.
\newblock Journal of Statistical Software. 2008;25(1):1--18.
\newblock doi:{10.18637/jss.v025.i01}.

\bibitem{factoextra}
Kassambara A, Mundt F. factoextra: Extract and Visualize the Results of Multivariate Data Analyses; 2020.
\newblock Available from: \url{https://CRAN.R-project.org/package=factoextra}.

\end{thebibliography}
\newpage

\section*{Supporting information}
\renewcommand{\thefigure}{S\arabic{figure}}
\setcounter{figure}{0}

\begin{figure}[!ht]
    \caption{{\bf Reaction distributions to news posts across countries.}}
    \label{fig:engagement_si_n}
    \includegraphics[width = .9\linewidth]{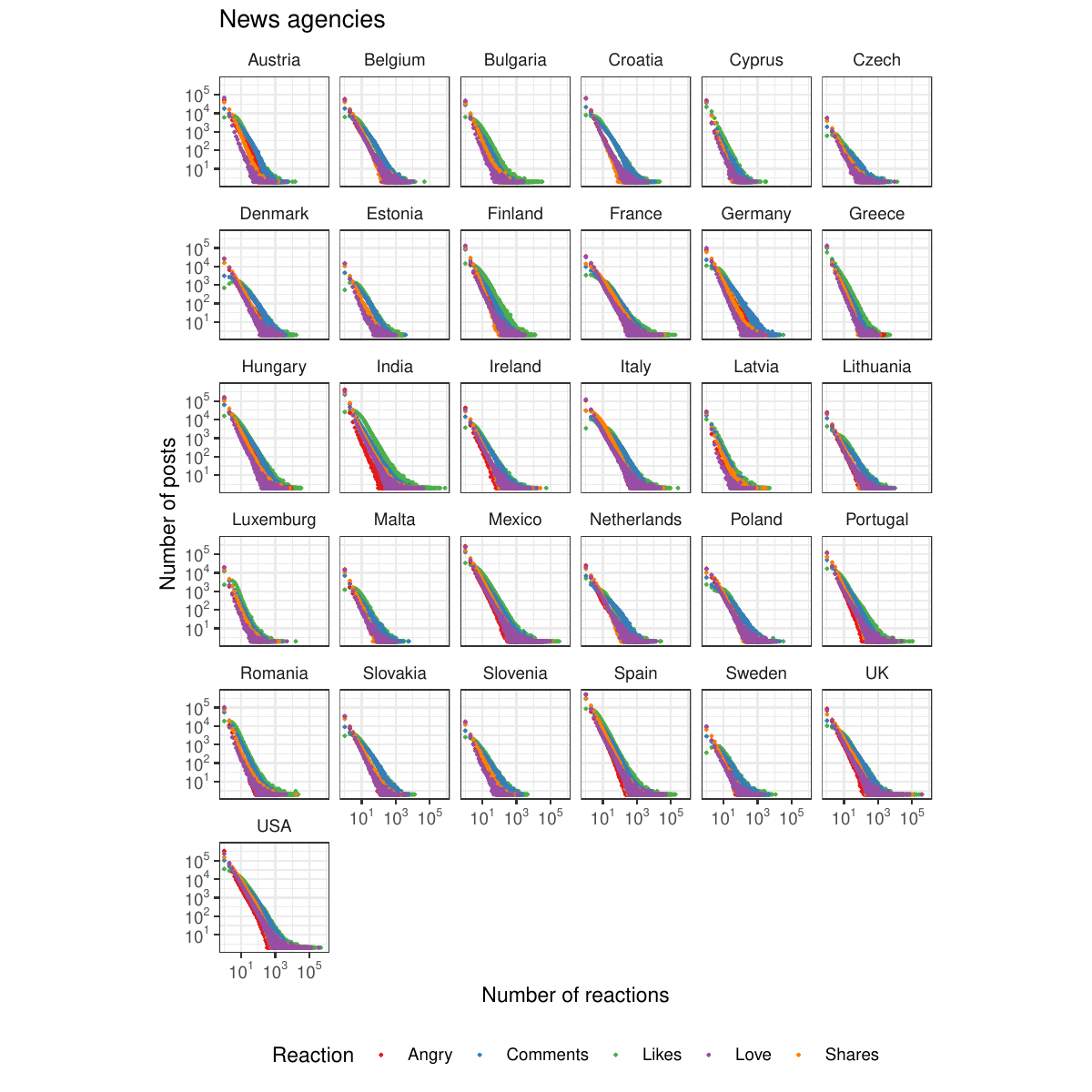}
\end{figure}

\begin{figure}
    \caption{{\bf Reaction distributions to posts from political parties across countries.}}
    \label{fig:engagement_si_p}
    \includegraphics[width = .9\linewidth]{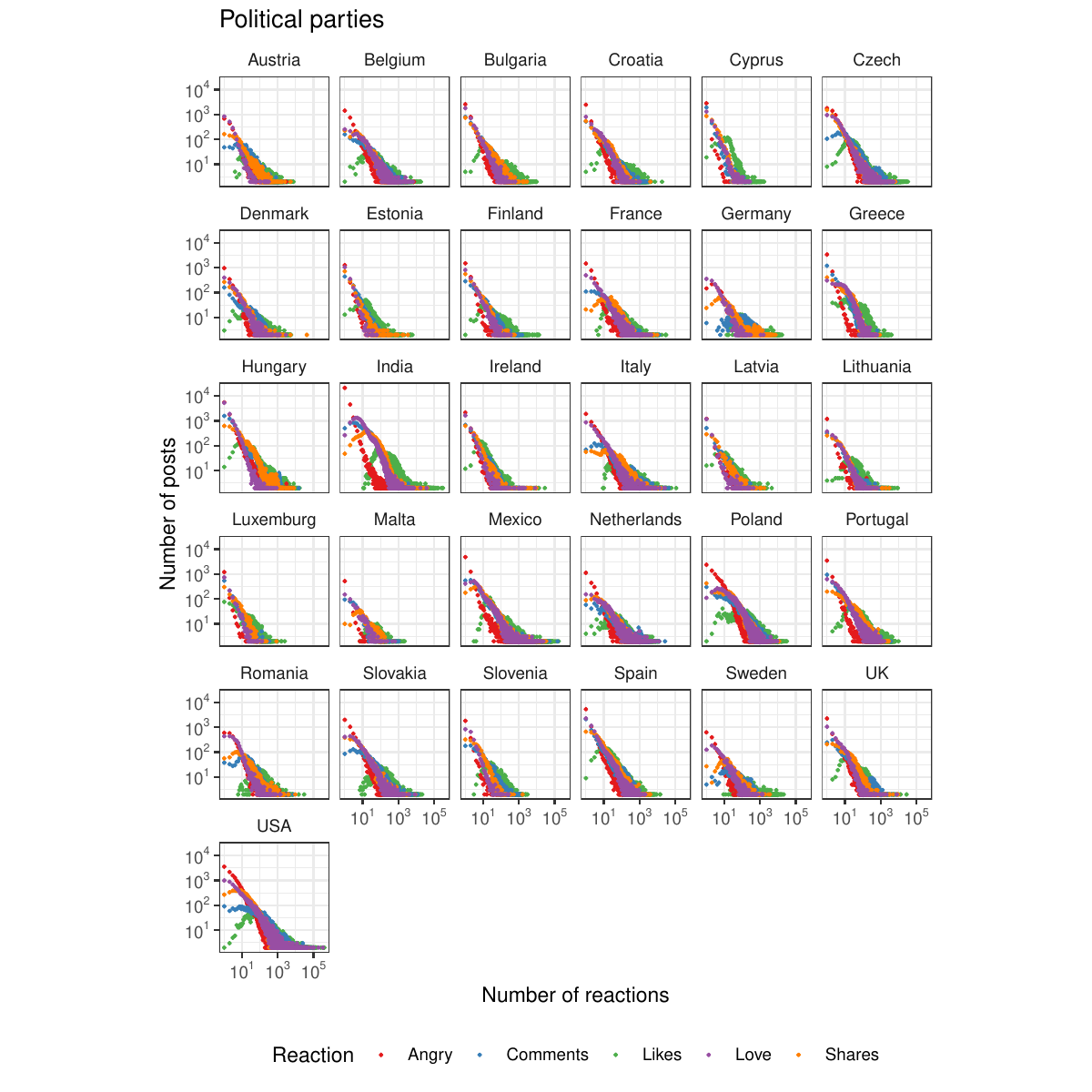}
\end{figure}

\begin{figure}
    \centering
    \includegraphics[width=\linewidth]{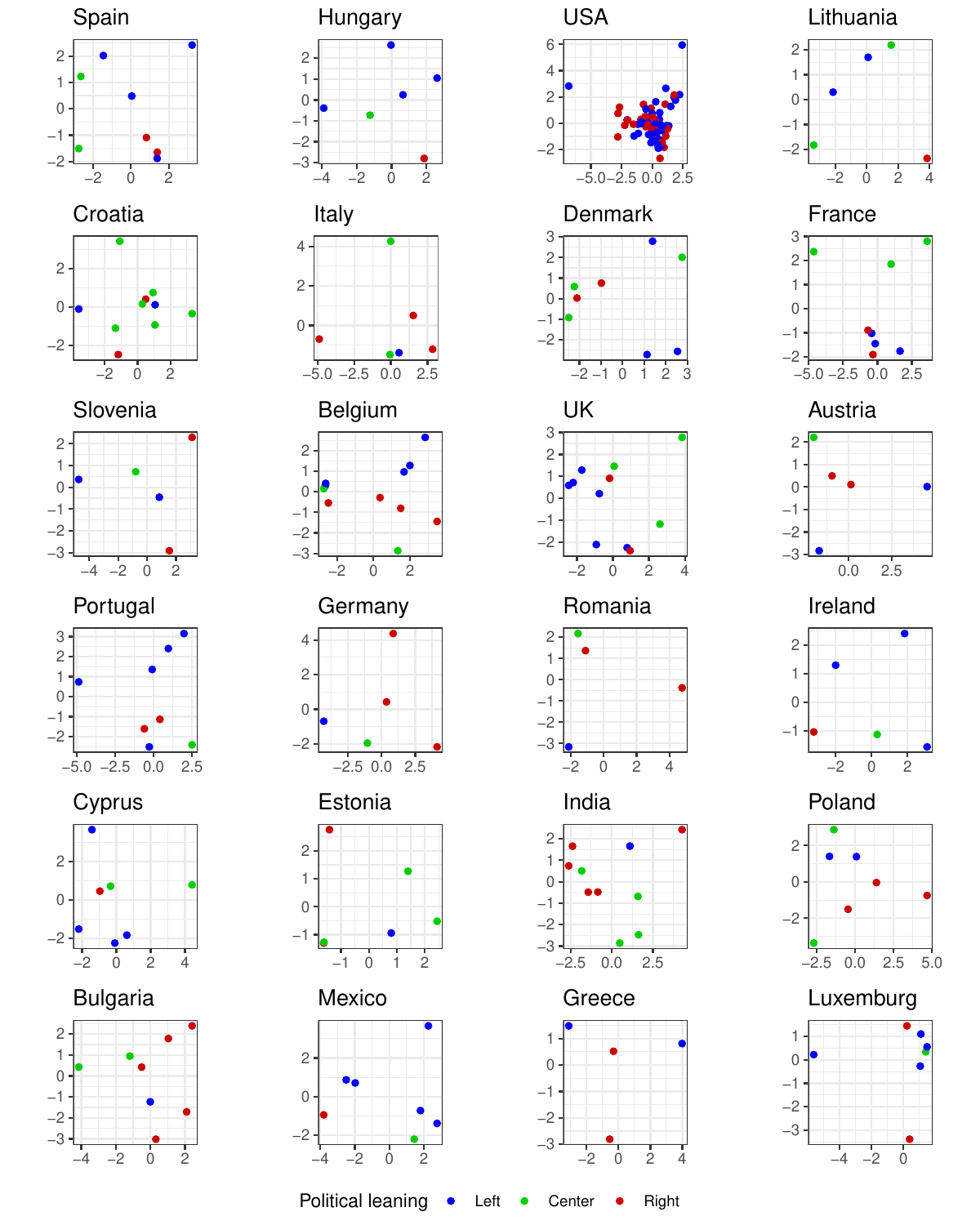}
    \caption{\textbf{Biplot of parties grouped by engaged topics.}}
    \label{fig:pca_parties_topics}
\end{figure}

\begin{figure}
    \caption{{\bf More engaging topics for parties.}}
    \label{fig:pca_parties_leanings_si}
    \includegraphics[width = .9\linewidth]{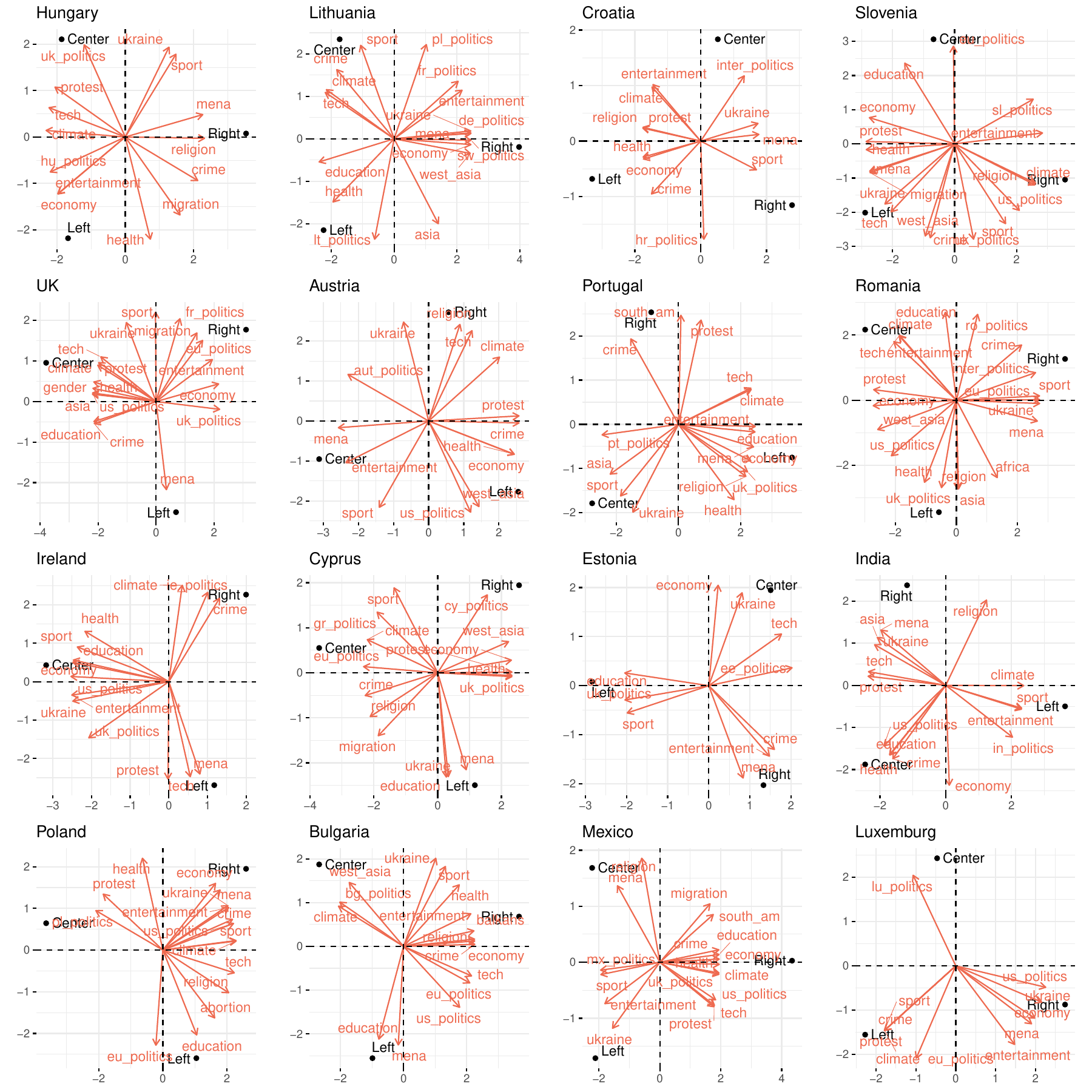}
\end{figure}







\end{document}